# Fast Trace Generation of Many-Core Embedded Systems with Native Simulation

David Castells-Rufas, Jordi Carrabina
Universitat Autònoma de Barcelona
Edifici Enginyeria, Campus UAB
08193 Bellaterra, Spain
{david.castells,jordi.carrabina}@uab.es

Pablo González de Aledo Marugán Pablo Sánchez Espeso
University of Cantabria
Avda de los Castros S/N
39005 Santander, Spain
{pabloga,javiergb,sanchez}@teisa.unican.es

## ABSTRACT

Embedded Software development and optimization are complex tasks. Late availably of hardware platforms, their usual low visibility and controllability, and their limiting resource constraints makes early performance estimation an attractive option instead of using the final execution platform.

With early performance estimation, software development can progress although the real hardware is not yet available or it is too complex to interact with.

In this paper, we present how the native simulation framework SCoPE is extended to generate OTF trace files. Those trace files can be later visualized with trace visualization tools, which recently were only used to optimize HPC workloads in order to iterate in the development process.

## Categories and Subject Descriptors

B.8.2 **[Hardware]**: Performance and Reliability – *Performance Analysis and Design Aids*; C.4 **[Computer Systems Organization]**: Performance of Systems – *Measurement Techniques*

## 1. INTRODUCTION

The design of high performance embedded systems is a challenging task. Its high complexity comes from the high number of dimensions of the design space, but also from the low visibility and controllability inherent to this kind of systems.

Several approaches help to mitigate the hurdles found on the different phases of design, like Hardware synthesis from high level programming languages, model based design descriptions, or, more often, the reuse of large complex blocks like microprocessors to form heterogeneous or homogeneous multi and many-processors.

As the number of cores per system is following an increasing trend, the verification of such systems becomes a hard issue. It is often not dominated by hardware verification (either at RTL or TLM level) but by parallel or concurrent software verification.

As performance optimization plays an important part of the verification process it is important to start it as soon as possible to be able to iterate faster in the development process (as shown in Figure 1). It is desirable to have a very fast simulation of a time accurate system, but there is a tradeoff between time accuracy and simulation speed. A summary of the different methods to get performance information is presented in [1].

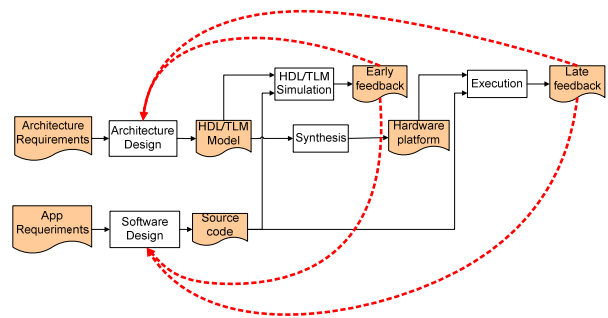

**Figure 1 Development process for complex Embedded Systems**

Early performance estimation in virtual platforms is usually limited to reporting total task time, or function profiling information, at most. In this paper, we present a new fast method to generate application trace logs based on native simulation that allows getting a very early and detailed feedback of the system performance. This approach allows analyzing the system software performance with some time accuracy even if the hardware system is not fully available.

The remainder of this paper is organized as follows. In the following section we will review how trace-based performance analysis technique is generally applied in HPC workloads. In section III we present the SCoPE native simulation framework. Its extension to generate trace files is presented in section IV. In section V we give some results obtained by analyzing some selected applications, and we finally end with conclusions.

## 2. TRACE-BASED PERFORMANCE ANALYSIS

Post-mortem trace analysis is a usual technique used by the HPC community to optimize parallel applications. Parallel applications are usually derived from an initial sequential source code. The parallelization process usually starts with a serial application that is analyzed, either by hand or semi-automatically, to detect the regions of the code that contain potential parallelism. A fairly easy way of parallelizing is by using OpenMP compiler directives. With this approach the compiler takes care of managing the threads needed for the code to work in parallel.

Tools exist to address the need of identifying computational and communication bottlenecks. A usual approach is to instrument the application to produce traces that can be later analyzed to find those bottlenecks. Tools like Vampirtrace [2] and Vampir [3][4], Paraver [5], TAU [6] are commonly using this approach to perform performance analysis on very-large systems with thousands of processors.

Instrumentation can be performed in two ways: either manually or automatically. Automatic instrumentation is done by using compiler techniques to introduce hooks at the enter and the leave sections of each function without programmer intervention. The hooks are usually directed to a function to produce a log message in the trace file (see Figure 2). With this technique one can analyze the amount of time that the processor has been running inside each function or the number of times each function is called.

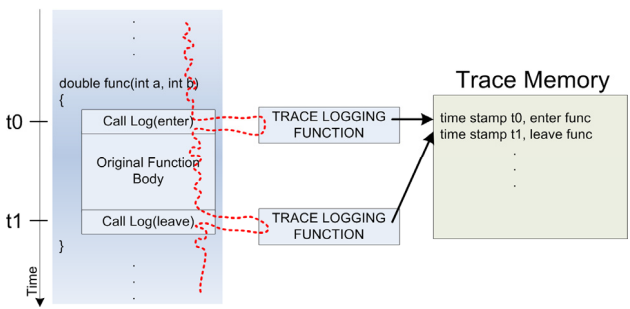

**Figure 2 Tracing Method**

These tools give much richer information than a simple profiler because they add time dynamics and therefore preserve spatial and temporal behavior of the running program. Furthermore, with proper instrumentation of communication functions, one can determine the communication pattern and overhead incurred by the application.

This is very useful, but there are several drawbacks mainly motivated by the instrumentation of the original application. When the application is instrumented, a small number of instructions are added to produce the log. Logs are usually written in memory first to minimize the time spent in slow disk access. Afterwards, when the analysis session is finished or the memory buffers have been filled, the logs are flushed to disk.

With complex functions, which are called at a low frequency, the overhead introduced by the instrumentation is small compared with the time spent inside the function. However, for small functions the instrumentation time is comparable or even higher than the time inside the function and so the instrumentation entirely modifies the program behavior.

Modern Object Oriented (OO) languages encourage the programmer to use get/set methods, and pass the responsibility to the compiler to optimize the application performance by function inlining when possible. Using the function instrumentation provided by the compiler will then switch off all optimizations for these functions regardless of the chosen optimization level (Intel compiler version>10 and GNU compiler) which renders the tracing information useless. IBM and PGI compiler will favor inlining over instrumenting – with these compilers therefore all the trace information is not available at all. The usual alternative from the HPC community is to go to manual instrumentation. With manual instrumentation the programmer tells which functions he is interested in, while being careful to avoid tracing high frequently called functions. However when using this approach one can lose important information about the behavior of those non instrumented functions.

So, the dilemma is to choose between 1) losing visibility by instrumenting just a subset of the functions and 2) analyzing a completely altered behavior because of the excessive instrumentation overhead.

## 3. NATIVE SIMULATION

In this work, a virtual platform based on native simulation [7][8] is used to provide fast and accurate performance estimation of many-core HW/SW systems. In native simulation, the application source code is annotated with performance-oriented code that is target-platform dependent [8]. During execution, the annotated code enables the estimation of the power consumption and application-code execution time in the target many-core platform.

The annotated code is compiled with a native compiler of the host computer in which the simulation (annotated code execution) is performed. These platforms allow the embedded software development process to be started even before the HW platform is completely defined because a limited number of high-level HW-platform parameters are needed. In the current version of the tool, it is possible to model the following elements of many-core architectures:

• Processor: it is characterized by defining the number of cycles that each instruction consumes.

• Bus interconnection: it is modeled by a SystemC generic bus model.

• Memories: the model of the memory hierarchy parameterizes memory response delay and memory size. Both data and instruction caches can be simulated.

• NoC: The tool provides two simulation approaches for the NoC. The first approach provides a high-level SystemC model where NoC transactions are modeled as blocks of data that flow from one node to another towards a virtual path.

The second approach provides a low-level cycle-accurate simulation in which all the micro-architectural details of the router crossbar and switches are modeled. A unique network interface is provided for both approaches and a designer can use the former (if fast simulations are the main goal) or the latter (if simulation accuracy is the main goal) approach.

### 3.1 Source Code Timing Annotation

A central part of the estimation technique is the annotation of estimated execution time of the software. Figure 3 shows the main steps of the process which analyzes directly the C source. First the code is characterized at compile time, and then timing estimations are used at execution time.

The application C/C++ source-code is analyzed by a parser to obtain a language-independent representation (XML model). In contrast to other basic-block-based approaches for performance and time estimation, the annotation is done directly in the C-source code, so the model execution is fast even for big and complex systems.

The annotation process has two steps. During the first step (characterization), some assembler directives are inserted into the code (see Figure 4). The goal is to directly identify the basic-block code even with compiler optimization.

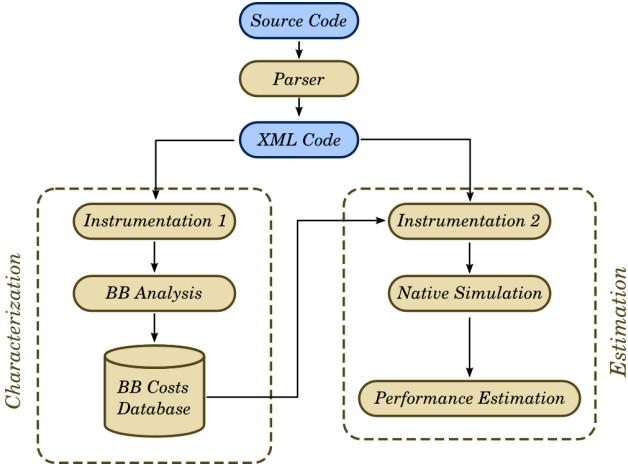

**Figure 3 Timing estimation process of Source Code**

The C-annotated code is then compiled with a target compiler and the resultant code is analyzed to detect the marks introduced. The output of this process is a database that characterizes each basic block, storing its number of instructions.

| Original Code | Annotated Code |
|---|---|
| `if (a[0] == 0)`<br>`{`<br>`  a[0] = 1;`<br>`}` | `asm("b _uc_mark_2_am3__");`<br>`if (a[0] == 0) {`<br>`  asm("b _uc_mark_3_rm__");`<br>`  a[0] = 1;`<br>`  asm("b _uc_mark_4__");`<br>`}`<br>`{asm("b _uc_mark_5_ai1__");}` |

**Figure 4 Annotated code**

During the second step (native simulation, right-side of Figure 3), the application code is annotated with a pair of functions per basic block that provide power and performance estimation. This functions use two additional parameters that are generated during host execution. The description of the techniques that generate these parameters are out of the scope of the paper (the real host machine caches are used to perform the estimation so the inherent hardware parallelism has not to be simulated). These parameters compute the number of cache misses in a basic block:

- *ICmisses* : number of instruction cache misses
- *DCmisses*: number of data cache misses

The description of the techniques that generate these parameters are out of the scope of the paper [15]. With this annotation and these parameters, the execution time of a block can be estimated as:

$$T = C \cdot T_m + T_{imiss} \cdot ICmisses + T_{dmiss} \cdot DCmisses$$

Being C the number of instructions of the basic-block, $T_m$ the mean time per instruction, $T_{imiss}$ the mean time spent per miss of the instructions cache and $T_{dmiss}$ the mean time spent per miss in the data cache

The time needed to execute an instruction is considered constant and the same for all instructions. In reality, though, there are differences in time costs between different instructions in the ISA and also among the same instructions executed at different times.

The first approach to characterize an instruction, in terms of execution time, is by assigning a constant value that is added during estimation time for each instruction execution. This instruction cost can be thought as the cost associated with the basic processing needed to finish the instruction. However, this base cost is also affected by other inter-instruction effects that can occur in real programs, like buffer prefetches, write buffer stalls, pipeline stalls, and cache misses. Base cost per instruction does not take in consideration the impact of these effects, so separate costs need to be assigned to them.

However, extra cost depends on input data sets of the system so this information is not available at compilation time, as the input data is unknown in this step. As the annotation process is made in the compilation step and the cost in terms of instructions is not known at this time, assigning a mean cost in this step implies a lose of accuracy.

### 3.2 Parallelism modelling in SCoPE

In order to create parallel tasks and spread them over different cores, SCoPE relies on the OpenMP parallelization methodology. As it can be observed in Figure 5, the parallelization of OpenMP annotated code lies over three different APIs. The OpenMP `pragma` directives are converted by gcc's front-end into a set of calls to functions defined by the `LIBGomp` library.

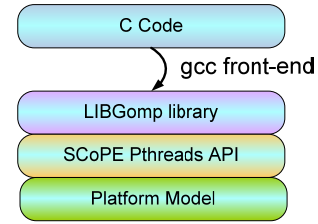

**Figure 5 APIs involved in OpenMP parallelization and estimation**

To model the effect of thread parallelism with SCoPE, the Linux API should be modified and a model of the operating system has to be included in that library. A modified version of the OpenMP library provided with SCoPE keeps track of the execution time needed for each task enabling complete time estimation.

As already mentioned, one of the most important aspects to consider when creating applications for many-core platforms is the efficient use of the hierarchical memory levels. This is an important aspect to consider. However, there is not a standardized API to manage DMA transfers yet, or to assign specific data localization for variables. The approach used in the literature uses specific "ad-hoc" functions or `pragma` directives that are called when the data transfers are performed. The problem of this approach is that these APIs are very different from platforms. The approach considered in SCoPE relies on "shared" and "private" `pragma` directives of OpenMP to assign a location in the hierarchical memory for every variable so this access time can be estimated. This provides a generic and practical way to simulate the code in different platforms. This also facilitates the native-simulation approach, as these `pragma` directives do not interfere with the usual functionality of the programs. To accomplish this, the parser of the SCoPE compiler has been improved to extract information about the OpenMP `pragma` directives.

## 4. TRACE GENERATION

As stated in [9] there are several benefits from generating traces from a virtual platform. First, the storage space and technology is not limited to the means of the target platform. So, large RAM memories and disks available in the host can save large trace files that can easily grow up to the order of Megabytes.

Second, the software can be developed and optimized without the need of building or distributing the hardware platform. This is an important issue because to the design time of the SoC, one usually has to add the time to develop and test a design kit based on it, and the number of initially available design kits is usually scarce.

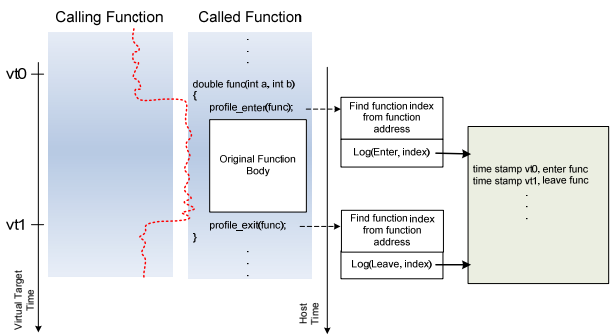

**Figure 6 The execution time is not modified from the target's time perspective when trace generation is enabled.**

Third, the simulation speed can be even faster than the final execution speed. This could be possible, for instance, if the target system is running at a clock frequency in the order of MHz while the host system is running at the order of GHz.

Finally the trace generation does not consume cycles of the target system (see Figure 6). In this sense trace generation in virtual platforms is a non-invasive method of observation that does not modify the execution time characteristics.

Trace generation can be either be manual or automatic. Manual instrumentation is done when the developers manually add the logging function invocation at the prolog and epilog of the functions of their interest. This method is interesting if there is no need to capture all the details from every function invocation, as capturing less events can led to important storage savings.

On the other hand automatic instrumentation can be used if all the function invocations should be stored. Gcc supports automatic function instrumentation with the flag `-finstrument-functions`. If this flag is enabled the compiler expects a couple of functions that are invoked from the prolog and epilog of every compiled function.

In either case a logging function is responsible to write a log in a standard trace format like OTF (Open Trace Format [10]) . In order to analyze the results as part of the iterative development process, the OTF traces are later analyzed in trace visualization tools like Vampir [3][4].

## 5. RESULTS

To test the system we use two applications. The first one is n-queens, a simple application which computes the different solutions of placing a number of n queens in a chess board of n by n tiles in such a way that they not threaten each other.

This application is not a typical embedded workload but it is convenient because it can be simply parallelized by using simple OpenMP `pragma` directives and does not require accessing external files through I/O interfaces. (See source code listing in Figure 8).

Since SCoPE supports parallel OpenMP workloads, as described in section III, the application can be directly executed in the native simulation environment.

We test the execution of the application with n=5 in different virtual platforms with different number of cores. The used cores are equivalent to arm926tnc as they support its instruction set. Trace generation adds a small overhead when flushing trace data to disk, but in all virtual platforms the execution of the application takes less than one second of the host time. In terms of target time the application execution time depends on the virtual platform in use. For instance in 16-core platform it takes 27 ms to execute. But this time is no effected at all by the time devoted to trace logging since this only consumes host time.

The benefit of generating traces is that performance problems can be better analyzed with visualization tools and better decisions can be taken to improve the code.

Figure 7 shows how trace visualization tools present the collected information. On the top right hand side of the picture we can find a report of the execution time by application function. This is the kind of information that we could get from a profiler. The left top panel shows the dynamics of the application and all its threads. Below we can show how the call-stack of each thread is progressing. In the example we just show the call-stack from processor 10 and 13.

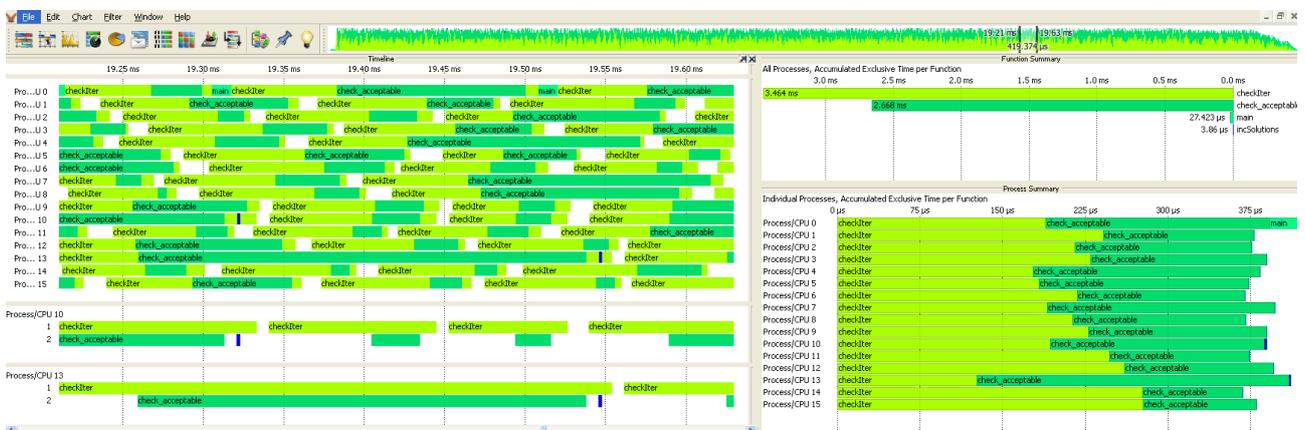

**Figure 7 Trace visualization of n-queens application on a 16-core many-core processor virtual platform**

Looking the source code of Figure 8 one could suspect that the `pragma omp critical` directive might impact negatively the performance of the application. In fact, many n-queens implementations use a reduction construct to avoid the apparent bottleneck. However the visualization of the application dynamics makes clear that this is not the case, since the increment of the `number_solutions` variable (in blue in Figure 8) is very seldom invoked.

```
int main(int argc, char* argv[]) {
   ...
   #pragma omp parallel for
   for (iter = 0; iter < max_iter; iter++) {
      int code = iter;
      int queen_rows[MAX_N];

      for (i = 0; i < n; i++) {
         queen_rows[i] = code % n;
         code /= n;
      }

      if (check_acceptable(queen_rows, n)) {
         #pragma omp critical
         number_solutions++;
      }
   }
}
```

**Figure 8 OpenMP version of the n-queens problem**

The second application analyzed is a JPEG encoder, a typical embedded workload. There are several implementations of the image compression algorithm. We have selected the source code of the project JPEGANT (see [11]) as a base because its specially addressed to embedded systems. The code has been parallelized following the first strategy presented in [12] which was based on [13] but is not specifically address to embedded systems. The pseudocode of the implementation is shown in Figure 9. Some operations like color conversion, DCT transform, and quantization can be performed in parallel since there is no data dependency among different blocks of the image. However huffman encoding must be serialized because of several data dependencies of the algorithm.

As in the previous example, different Hardware configurations are tested with the same application. The visualization of the traces (as shown in Figure 10) give a clear idea of why the application does not scale well above a certain number of cores (typically 5 as explained in [12]). After the parallelization of the first loop the sequential part of the algorithm (huffman coding, and I/O) dominates.

**Table 1. Execution time for JPEG on 16-core virtual platform**

|  | Host | Target |
|---|---|---|
| Platform | Intel XEON E5620 | virtual arm926tnc |
| Cores | 4 | 16 |
| Clock Frequency | 2.40 GHz | 470 Mhz |
| Excution Time | 18ms | 26ms |

In this case, the simulation of the application running on a 16-core target system is executed in 18ms, in hosts system time (see Table 1). So, as we mentioned before target expected execution time is bigger than the simulation time, even with the overhead of trace generation.

```
int main (int argc, char *argv[]) {
   ...
   #pragma omp parallel for
   for (int iter=0; iter < maxIter; iter++) {
      ...
      getBlock(x, y, 16, 16, (BGR*)RGB16x16));
      ...
      subsample2(RGB16x16, Y8x8, Cb8x8, Cr8x8);
      ...
      dct3(Y8x8[0][0], HY8x8[iter][0][0]);
      dct3(Y8x8[0][1], HY8x8[iter][0][1]);
      dct3(Y8x8[1][0], HY8x8[iter][1][0]);
      dct3(Y8x8[1][1], HY8x8[iter][1][1]);
      dct3(Cb8x8, HCb8x8[iter]);
      dct3(Cr8x8, HCr8x8[iter]);
   }
   for (int iter=0; iter < maxIter; iter++) {
      huffman_encode(HUFFMAN_CTX_Y, HY8x8[iter][0][0]);
      huffman_encode(CTX_Y, HY8x8[iter][0][1]);
      huffman_encode(CTX_Y, HY8x8[iter][1][0]);
      huffman_encode(CTX_Y, HY8x8[iter][1][1]);
      huffman_encode(CTX_Cb, HCb8x8[iter]);
      huffman_encode(CTX_Cr, HCr8x8[iter]);
   }
   ...
}
```

**Figure 9 OpenMP version of the JPEG encoder problem**

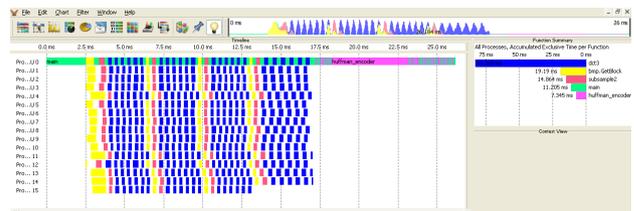

**Figure 10 Visualization of the traces generated for the JPEG application in a 16-core many-core**

## 6. CONCLUSIONS

Fast trace generation from parallel workloads running on multi and many-core embedded systems is possible with native simulation. Native simulation provides a fast early estimation of application expected performance and can be applied to work on parallel software bottlenecks even before the hardware platform is available.

Besides avoiding the typical limitations of embedded target platforms, we have shown how native simulation provides several important benefits.

First the storage space is not limited to that of the target platform. But, more importantly, the speed of the simulation together with trace generation can be even faster than execution in the final platform. Moreover, with this method trace generation consumes no virtual target time. So, compared with trace generation from the execution platform, this method is better because it does not modifiy the time characteristics of the application under test.

## ACKNOWLEDGMENTS
This work was partly supported by the European cooperative ITEA2 projects 09011 H4H and 10021 MANY, the CATRENE project CA112 HARP, the Spanish Ministerio de Economía y Competitividad project IPT-2012-0847-430000, the Spanish Ministerio de Industria, Turismo y Comercio projects TSI-020100-2010-1036, and TSI-020400-2010-120, and the Spanish Ministerio de Ciencia e Innovación project TEC2011-28666-C04-02 .

## REFERENCES

[1] Castells-Rufas, D.; Fernandez-Alonso, E. & Carrabina, J. "Performance Analysis Techniques for Multi-Soft-Core and Many-Soft-Core Systems International" *Journal of Reconfigurable Computing, 2012*, 2012

[2] T. William, H. Mix, B. Mohr, F. Voigtländer, R. Menzel "Enhanced Performance Analysis of Multi-Core Applications with an Integrated Tool-Chain" *In Proceedings Parallel Computing 2009 (PARCO),* Lyon, 2009

[3] W. E. Nagel, A. Arnold, M. Weber, H.-C. Hoppe, and K. Solchenbach, "VAMPIR: Visualization and Analysis of MPI Resources," *Supercomputer, vol. 12*, pp. 69–80, 1996.

[4] H. Brunst, H.-C. Hoppe, W. E. Nagel, and M. Winkler, "Performance Optimization for Large Scale Computing: The Scalable VAMPIR Approach," pp. 751–760, 2001.

[5] V. Pillet, J. Labarta, T. Corts, and S. Girona, "PARAVER: A Tool to Visualize and Analyse Parallel Code," in *Proceedings of WoTUG-18: Transputer and occam Developments, volume 44 of Transputer and Occam Engineering*, 1995, pp. 17–31.

[6] R. Bell, A. Malony, and S. Shende. "A Portable, Extensible, and Scalable Tool for Parallel Performance Profile Analysis". *In European Conference on Parallel Processing (EuroPar), volume LNCS 2790*, pages 17–26, September 2003.

[7] Posadas, H.; Real, S. & Villar, E. "M3-SCoPE: Performance Modeling of Multi-Processor Embedded Systems for Fast Design Space Exploration Multi-Objective Design Space Exploration of Multiprocessor SOC Architectures: The Multicube Approach", *Springer Verlag, 2011*, 19

[8] Marugán, P.G.D.A., González-Bayón, J., Espeso, P.S. "A virtual platform for performance estimation of many-core implementations" *(2012) Proceedings - 15th Euromicro Conference on Digital System Design, DSD 2012*, art. no. 6386939, pp. 541-544.

[9] Castells-Rufas, D.; Joven, J.; Risueño, S.; Fernandez, E.; Carrabina, J.; William, T. & Mix, H. "MPSoC performance analysis with virtual prototyping platforms" *Proceedings of the International Conference on Parallel Processing Workshops, 2010*, 154-160

[10] A. Knüpfer, R. Brendel, H. Brunst, H. Mix, W. E. Nagel: Introducing the Open Trace Format (OTF) In Vassil N. Alexandrov, Geert Dick van Albada, Peter M. A. Sloot, Jack Dongarra (Eds): *Computational Science - ICCS 2006: 6th International Conference, Reading, UK, May 28-31, 2006, Proceedings, Part II, Springer Verlag*, ISBN: 3-540-34381-4, pages 526-533, Vol. 3992, 2006.

[11] JPEGANT https://code.google.com/p/jpegant/

[12] Castells-Rufas, D.; Joven, J. & Carrabina, J. "Scalability of a Parallel JPEG Encoder on Shared Memory Architectures" *2010 39th International Conference on Parallel Processing*, 2010, 502-507

[13] PJPEGENC https://code.google.com/p/pjpegenc/